\documentclass{INTERSPEECH2023}

 \interspeechcameraready

\usepackage{amsmath}
\usepackage{adjustbox, booktabs}

\title{Speech Self-Supervised Representation Benchmarking: \\ Are We Doing it Right?}

\name{Salah Zaiem$^1$, Youcef Kemiche$^{2,3}$, Titouan Parcollet$^{4,5}$, Slim Essid$^1$, Mirco Ravanelli$^6$}
\address{$^1$LTCI, Télécom Paris, Institut Polytechnique de Paris, France\\
  $^2$Hi! PARIS Engineering Team, France \\
  $^3$Capgemini, France \\
  $^4$Samsung AI Center, Cambridge, United-Kingdom\\
  $^5$University of Cambridge, United-Kingdom\\
  $^6$ Mila-Quebec AI Institute, Université de Montréal, Concordia University, Canada}
\email{salah.zaiem@telecom-paris.fr}
\begin{document}
\maketitle
 
\begin{abstract}
Self-supervised learning (SSL) has recently allowed leveraging large datasets of unlabeled speech signals to reach impressive performance on speech tasks using only small amounts of annotated data. The high number of proposed approaches fostered the need and rise of extended benchmarks that evaluate their performance on a set of downstream tasks exploring various aspects of the speech signal. However, and while the number of considered tasks has been growing, most rely upon a single decoding architecture that maps the frozen SSL representations to the downstream labels. This work investigates the robustness of such benchmarking results to changes in the decoder architecture. Interestingly, it appears that varying the architecture of the downstream decoder leads to significant variations in the leaderboards of most tasks. Concerningly, our study reveals that benchmarking using limited decoders may cause a counterproductive increase in the sizes of the developed SSL models.\end{abstract}

\section{Introduction}
Self-supervised learning (SSL) provides a powerful means of benefiting from large volumes of unlabeled data to achieve substantial performance gains across a range of downstream tasks, without relying on manual annotations. Several approaches have been proposed in the literature, including predictive coding  \cite{baevski2020wav2vec}, multi-task learning \cite{zaiem2022pretext}, and contrastive learning approaches \cite{cola, zaiem2022automatic}. Self-supervised models have lately become a necessary tool for speech practitioners suffering from a lack of annotations in a growing set of tasks. 

However, experimenting with large SSL models is a costly endeavor both in terms of time and computing. Thus, the proliferation of approaches for speech SSL \cite{Mohamed2022} has highlighted the need for benchmarks that evaluate their performance across multiple downstream tasks. Ultimately, such benchmarks explore various aspects of the speech signal, helping practitioners to make informed choices adapted to their use cases. They also serve as a simple way to assess the effectiveness of different techniques and identify areas for improvement. As a result, an increasing number of extended benchmarks have emerged in recent years, providing standardized frameworks for evaluating the performance of speech SSL models and algorithms \cite{superb, superb_slt, evain2021lebenchmark}. These benchmarks cover a large range of speech tasks and applications and, even among a single task such as automatic speech recognition (ASR), a large choice of linguistic, acoustic and prosodic settings \cite{tsai-etal-2022-superb}. 

In popular speech SSL benchmarks, for every considered task, the self-supervised representation is probed using a learned downstream decoder mapping the frozen representation to the final labels. Simplicity and limited capacities are the main drivers of the choice of these downstream probes with, for instance, linear probing for classification tasks or shallow vanilla recurrent neural networks for speech recognition. We hypothesize that such a benchmarking policy may harm the development of novel SSL technologies in at least two ways. First, the popularity of the main benchmarks, \textit{e.g.} SUPERB \cite{superb}, has made the considered downstream probes the \textit{de facto} evaluation setting for any new speech SSL model. This is also true for the metrics used in the benchmark, which will condition the optimization process during the development of these new approaches, possibly leading to decisions to drop models that perform poorly with the selected probes but may perform better with other downstream architectures. Second, as the simplicity of the probes contrasts with the growing complexity of the SSL encoders, testing with low-capacity probes may imply a transfer of complexity from the decoder (expected to be task-specific) to the encoder (supposedly more general), leading to superfluously large self-supervised parts. As an example, in computer vision, Dubois \textit{et al.} \cite{dubois2022improving} have shown that changing the probe family from linear to multi-layer perceptrons (MLP) leads to different optimal choices in the hyperparameters of the SSL models and enables smaller SSL representations.

A first solution to such limitations might come from considering novel downstream-agnostic evaluations \cite{problemagnostic}, through intrinsic quality assessment metrics for speech embeddings \cite{schatz}, but the correlation of these metrics with downstream performances has not been clearly identified \cite{algayres2020evaluating} yet. In the context of image classification, Garrido \textit{et al.} \cite{garrido2022rankme} have indicated that the rank of the vision SSL representations correlates highly with the final downstream performance. However, this study has been conducted with linear downstream probes only.  Conscious of these issues, SUPERB \cite{superb}, the main speech SSL benchmark, proposes two tracks where the downstream probes can be chosen by the model submitter, with or without capacity constraints on the probing architectures. However, to this day, these two tracks remain empty of submissions. 

Therefore, this work evaluates the robustness of speech SSL benchmarks to variations in the probing architecture. The contributions are three-fold: i) We benchmark a set of published state-of-the-art SSL models on various speech tasks, varying the downstream decoders (Section \ref{sec:setting}); ii) We show that, except for ASR on LibriSpeech, the rankings and relative performances of the models are severely impacted by a change of decoder (Section \ref{sec:results}); iii) We release the code\footnote{\url{github.com/salah-zaiem/speechbrain-2/tree/develop/recipes/SSL_benchmark}} for the benchmark, implemented in SpeechBrain \cite{speechbrain}, for replication and further research on SSL models and downstream tasks.

\section{Definition and Experimental Protocol}
\label{sec:setting}
This section formally describes the limitation faced by current speech SSL benchmarks and also details the experiment protocol illustrating this issue.

\subsection{Problem definition}
Formally, a SSL pipeline consists of two models: a pre-trained encoder $\phi$ and a downstream probe $f$. $\phi$ is learned through solving a pretext task on unlabeled speech datasets (Libri-light \cite{Kahn_2020} and LibriSpeech960 \cite{libri} have been popular choices in the literature), while $f$ is learned on the annotated downstream dataset. In this framework, the SUPERB benchmark has chosen for every considered downstream task $T$ a probing family $\mathfrak{F}_T$ (\textit{i.e.} a downstream architecture) and shows for every considered SSL encoder $\phi$ a task error rate corresponding to: 
\begin{equation}
\label{eq:superb}
\min\limits_{f \in \mathfrak{F}_T} E_t(f\circ \phi)
\end{equation}
with $E_t(f\circ \phi)$ the test-set error rate of the full SSL pipeline. 

However, proper benchmarking would rank the candidate models according to the minimum quantity defined above over all the possible downstream probing families, possibly with a model capacity constraint. This measure is computationally intractable, as it would require training a model with every possible known downstream architecture respecting the capacity constraints for every considered encoder and task. In this work, we will explore whether benchmarking according to the value obtained in Equation \eqref{eq:superb} leads to a proper and robust ranking. To this effect, for every considered downstream task, we will survey two (or more) probing families, and see whether the rankings and relative differences obtained using the first families are kept in the second experiment.

\subsection{Self-supervised pretrained models} 

To perform this study, a subset of the models presented in the SUPERB benchmark is considered. In particular, we retrieved the best-performing and easiest-to-use SSL models.  Hence, $9$ models acting directly on the raw waveform have been selected: Wav2vec 2.0 \cite{baevski2020wav2vec}, HuBERT \cite{hubert}, WavLM\footnote{We considered WavLM Base+, trained on $94k$ hours of speech data, for the Base WavLM version.}\cite{Chen2021}, and Data2vec \cite{baevski2022data2vec} both in their Base and Large versions. DistilHuBERT \cite{Chang2021}, a distilled version of Hubert Base with $4$ times less transformer layers, is also added to the list. The considered models share the same output frequency  generating a representation of size $D$ every 20 ms of audio signal, with $D=1,024$ for Large versions and $D=768$ for Base ones and DistilHuBERT. 

All the models considered are based on very similar Transformer-based architectures, but their pretraining pretext tasks differ. Wav2vec2.0 training is based on the contrastive predictive coding \cite{oord2018representation} (CPC) objective, maximizing the mutual information between a set of context features and predicted future samples. HuBERT and WavLM learn to map the unlabeled audio to sequences of pseudo-labels obtained through the clustering of previous representations. WavLM introduces distortions during the training to enforce noise-related invariances. Inspired by teacher-student approaches, Data2vec is trained by utilizing a masked input view to predict latent representations of the complete input data, in a self-distillation configuration. All the pre-trained checkpoints are obtained from their HuggingFace (HF) official cards \cite{wolf2020transformers}, with the exception of Wav2vec2.0 Large whose Fairseq \cite{ott2019fairseq} checkpoint is used, as the HF one was underperforming compared to the results reported in SUPERB.

\subsection{Downstream Tasks and Datasets}
Speech SSL benchmarks have been attempting to evaluate universal speech representations by providing a large set of different tasks that assess different aspects of the speech signal. Similarly, we propose $7$ tasks that tackle the phonetic, speaker-identity-related, emotive and semantic aspects. \\

\noindent \textbf{Speech Recognition Tasks.} Four speech recognition tasks are considered in this work. For the two first ones, LibriSpeech \cite{libri} \textit{train-clean-100}/\textit{dev-clean} subsets are used for training and validation while \textit{test-clean} and \textit{test-other} are used for testing. To test the ability of the models with fewer data and in a spontaneous context, the Buckeye dataset \cite{buckeye} is considered. The training, validation, and test splits used in our Buckeye experiments are available in the repository with the training set containing approximately $9.5$ hours of audio and the test set $1.5$ hours. For these two English ASR tasks, two results are shown depending on whether a language model (LM) is used or not during decoding. In the experiments labeled as \textit{Without LM}, greedy decoding is applied. For the ``With LM" experiments, the LibriSpeech official $4$-gram language model is utilized. The language model is combined with shallow fusion to the acoustic model. Since low-resource languages are one of the main applications of SSL methods, two low-resource languages tasks, extracted from the CommonVoice $11.0$ \cite{ardila2020common} release, are considered: Welsh (Cymraeg) and Basque (Euskera). The Word Error Rate (WER) is used as the error metric for all the ASR tasks. In all the ASR experiments, the probe is trained using the Connectionist Temporal Classification\cite{ctc} loss at the character level. \\ 
\noindent \textbf{Automatic Speaker Verification.} Automatic Speaker Verification (ASV) involves a binary classification process to determine if the speakers in a pair of utterances are the same. As in the SUPERB benchmark, we use VoxCeleb1 \cite{Nagrani_2017} train and test splits for this task. It is important to note that the speakers in the testing set may not have been included in the training set. The evaluation metric is equal error rate (EER). \\
\noindent \textbf{Emotion Recognition (ER).} IEMOCAP \cite{busso2008iemocap}, a dataset containing $10,039$ from $10$ speakers, is used for Emotion Recognition. The task consists of predicting an emotion class for the utterance among $4$ candidates (neutral, happy, sad, angry). The performance shown is the mean of $10$ runs done cross-validating on ten folds leaving each time one of the speakers' data for testing. 
\begin{table*}[]
\caption{SSL benchmarking results for all tasks and downstream architectures. On most downstream tasks, rankings and relative performances are severely impacted by  the change of the downstream architecture. The number of parameters of the SSL encoder and the probes is shown in the ``Params" rows and columns. }

\centering
\scalebox{0.75}{
\begin{tabular}{lcccccccccccc} \toprule
\textbf{Models /Tasks} & \textbf{SSL Params.} & \multicolumn{4}{c}{\textbf{LibriSpeech train-100 ASR}} & \multicolumn{2}{c}{\textbf{Buckeye ASR}} & \textbf{Welsh} & \textbf{Basque} & \textbf{ASV} & \textbf{ER} & \textbf{IC} \\ \midrule
\multicolumn{2}{l}{Evaluation Metrics}        & \multicolumn{4}{c}{WER $\downarrow$} &      \multicolumn{2}{c}{WER $\downarrow$}                   & WER $\downarrow$           & WER $\downarrow$            & EER $\downarrow$     & Acc. $\uparrow$ & Acc. $\uparrow$ \\ \midrule

\multicolumn{2}{l}{First downstream architectures}        & \multicolumn{4}{c}{LSTM} &      \multicolumn{2}{c}{LSTM}                   & LSTM           & LSTM            & Xvectors     & Pool + Lin. & Pool + Lin. \\ \midrule

 &  & \textbf{Clean} & \textbf{Other} & \textbf{Clean LM} & \textbf{Other LM} & \textbf{w/o LM} & \textbf{with LM} & \textbf{Welsh} & \textbf{Basque} & \textbf{ASV} & \textbf{ER} & \textbf{IC} \\ \midrule
DistilHuBERT           & 23.5M               & 13.99              & 34.91               & 9.96                  & 28.26                 & 35.59            & 28.29               & 53.20          & 46.78           & 9.1        & 65&46.6                 \\
Wav2vec 2.0 Base       & 95M                 & 6.23               & 14.93               & 4.86                  & 11.97                 & 24.87            & 19.48               & 54.45          & 51.21           & 5.29         & 66.4   & 59.0           \\
Wav2vec 2.0 Large      & 317.4M              & 3.72               & 9.25                & 3.13                  & 7.48                 & \textbf{20.72}            & 16.11               & 45.42          & \textbf{37.98}           & 5.69         & 69.3 &66             \\
HuBERT Base            & 94.7M               & 6.24               & 15.03               & 5.03                 & 12.31                 & 45.53            & 26.51               & 52.92          & 46.91           & 4.50         &67.5 & 53.8                \\
HuBERT Large           & 316.6M              & 3.57               & 8.12                & 2.90                  & 6.59                  & 51.30            & 33.10               & 51.21          & 46.15           & 5.20         & 71.3 &69.9              \\
WavLM Base+            & 94.7M               & 5.96               & 14.33               & 4.84                  & 11.72                 & 42.21            & 24.41               & 51.31          & 46.40           & 3.74         & 67.1 & 57.9               \\
WavLM Large            & 316.6M              & 3.48               & 7.37                & 2.87                  & 5.96                  & 27.31            & \textbf{14.27}               & 48.92          & 41.89           & \textbf{2.98}    & \textbf{75.3}      & \textbf{78.8}               \\
Data2vec Base          & 93.8M               & 5.30               & 13.79               & 4.03                 & 10.97                 & 37.26            & 30.50               & 54.00          & 46.37           & 5.43        & 63.0    & 56.9             \\
Data2vec Large         & 314.3M              & \textbf{3.10}            & \textbf{6.50}            & \textbf{2.58}                & \textbf{5.38}                  & 22.63            & 18.63               & \textbf{44.32}          & 38.23           & 4.89      & 64.1   & 69.8                \\ \midrule
\multicolumn{3}{l}{\textbf{Probe size and inference metrics}} \\ \midrule
\multicolumn{2}{l}{Downstream Parameters Base}             & \multicolumn{4}{c}{39.9M}             & \multicolumn{2}{c}{39.9M}                & 40.3M          & 40.3M           & 7.0M         & 13.8k       & 3.1k        \\
\multicolumn{2}{l}{Downstream Parameters Large}               & \multicolumn{4}{c}{42M}                     & \multicolumn{2}{c}{42M}                  & 42.4M          & 42.4M           & 7.7M         & 18.4k       & 4.1k        \\ \midrule  \midrule

\multicolumn{2}{l}{Second downstream architectures}                     & \multicolumn{4}{c}{Conformer} &      \multicolumn{2}{c}{ContextNet}                   & Lin.           & Lin            &  ECAPA      & ECAPA & LSTM + Lin. \\ \midrule
 &  & \textbf{Clean} & \textbf{Other} & \textbf{Clean LM} & \textbf{Other LM} & \textbf{w/o LM} & \textbf{with LM} & \textbf{Welsh} & \textbf{Basque} & \textbf{ASV} & \textbf{ER} & \textbf{IC} \\ \midrule
DistilHuBERT                & 23.5M       & 14.97                              & 36.51                              & 11.54                 & 31.41                 & 58.56                         & 43.61               & 80.78               & 77.04                & 2.85          & 72.4       & 74.9        \\
Wav2vec 2.0 Base            & 95M         & 6.91                               & 15.39                              & 5.09                 & 12.29                 & 30.04                         & 23.04               & 74.31               & 71.76                & 2.82          & 73.2       & 77.7        \\
Wav2vec 2.0 Large           & 317.4M      & 4.32                              & 9.25                              & 3.58                  & 7.03                 & 23.92 & 18.68               & 75.45               & 78.48                & 3.17          & 68.4       & 79.0        \\
HuBERT Base                 & 94.7M       & 6.88                               & 15.68                              & 5.23                  & 12.63                 & 30.44                         & 23.11               & 77.39               & 73.40                & 2.40          & \textbf{78.2}       & 79.4        \\
HuBERT Large                & 316.6M      & 3.96                               & 8.60                               & \textbf{3.10}                  & 6.88                  & 39.39                         & 31.57               & 71.58               & 60.24                & 3.84          & 71.5       & 80.1        \\
WavLM Base+                 & 94.7M       & 6.55                               & 14.93                              & 4.98                  & 11.80                 & 27.73                         & 21.69               & 75.87               & 69.43                & \textbf{1.76}          & 72.6       & 81.2        \\
WavLM Large                 & 316.6M      & 4.08                              & 8.10                               & 3.13                  & \textbf{6.31}                  & \textbf{15.61}                         & \textbf{12.1}                & \textbf{68.73}               & \textbf{56.32}                & 1.77          & 77.4       & \textbf{85.8}        \\
Data2vec Base               & 93.8M       & 5.85                               & 14.32                              & 4.53                  & 12.52                 & 40.53                         & 33.45               & 77.49               & 75.26                & 3.75          & 72.0       & 73.4        \\
Data2vec Large              & 314.3M      & \textbf{3.43}                              & \textbf{6.82}                             & 3.27                  & 6.58                  & 25.26                         & 21.5                & 69.09               & 63.31                & 2.67          & 71.3       & 79.9        \\\midrule
\multicolumn{3}{l}{\textbf{Probe size and inference metrics}} \\ \midrule
\multicolumn{2}{l}{Downstream Parameters Base}                      &\multicolumn{4}{c}{11.2M}                     & \multicolumn{2}{c}{32.4M}                            & 1.9M                & 1.9M                 & 9.2M          & 7.3M       & 42M         \\
\multicolumn{2}{l}{Downstream Parameters Large}                       & \multicolumn{4}{c}{11.2M}                     & \multicolumn{2}{c}{32.5M}                        & 2.3M                & 2.3M                 & 9.8M          & 7.9M       & 44.1M       \\ \bottomrule  
\end{tabular}}
\label{tab:DS}
\end{table*}

\noindent \textbf{Intent Classification (IC)}. Instead of Speech Commands (SC) \cite{warden2018speech}, which is the dataset used in the SUPERB benchmarks for IC, we use the SLURP dataset \cite{bastianelli2020slurp}, as the error rates with SC are already very low and SLURP is a more challenging task. The SLURP collection consists of approximately $72,000$ audio recordings that capture single-turn user interactions with a home assistant. The IC task consists of classifying every utterance into one of the $18$ considered scenarios. Examples of scenarios include ``calendar", ``email" and ``alarm". The metric for emotion recognition and intent classification is accuracy. 

\subsection{Downstream Probes}
This section provides a high-level description of the used downstream probes. The numerous hyperparameters and architectural details, allowing thorough replication of the experiments, are available in the provided repository. \\

\noindent\textbf{Global settings.} During fine-tuning, the SSL encoder weights are frozen and only the downstream decoder weights are learned. Similarly to what is done in SUPERB, we observed that the last-layer representation may not always be optimal and, thus, collected all the hidden layers representations from the pretrained model. These hidden states are then weighted and summed together to create the representation fed to the decoder. The layer weights are trained during the fine-tuning. To check that the benchmark is properly set, we reproduced the SUPERB downstream architectures in the first set of experiments. Second, the probes are changed with simpler or more complex alternatives inspired from the relevant literature of every task. \\ 
\noindent \textbf{ASR tasks.} In the first set of experiments, reproducing the SUPERB conditions, a vanilla $2$-layer $1,024$-units BiLSTM followed by a linear layer mapping the audio to the characters are used. For the second downstream architectures, we consider an encoder-decoder Conformer \cite{gulati2020conformer} downstream architecture for LibriSpeech, with $12$ encoder layers, $4$ decoder ones and $4$ attention heads. For the Buckeye task, we use a ContextNet \cite{han2020contextnet} with unitary strides to keep the frame rate of the SSL models. For Welsh and Basque, a two-layered dense neural network is used to map every frame representation to the probabilities of the characters. In a second time, experiments using ContextNet with LibriSpeech are also performed. ContextNet and Conformer performances, which are close to state-of-the-art on LibriSpeech, motivated their selection as downstream probes. \\
\noindent \textbf{Automatic Speaker Verification.} First, the X-vector \cite{snyder2018x} architecture is tested with an AM-Softmax \cite{wang2018additive} loss training of the speaker embeddings. For verification, we use the cosine similarity between speaker representations. In the second experiment, we replace the X-vectors with an ECAPA-TDNN neural network \cite{desplanques2020ecapa}. ECAPA-TDNN combines time-delay neural networks with parallel attention mechanisms to capture temporal dependencies and the long-range context in sequential data and achieves state-of-the-art results in speaker verification. \\
\noindent \textbf{Classification tasks.} Similarly to SUPERB, we perform linear probing for classification tasks (\textit{i.e.} intent classification and emotion recognition) in the first set of experiments. The representations are first average-pooled along the time axis before being fed to a classification linear layer. For the second downstream architecture, and inspired by state-of-the-art approaches \cite{wang2021fine}, ECAPA-TDNN is chosen for emotion recognition. For intent classification, following published architectures \cite{lugosch2019speech},  two layers of BiLSTM with hidden size $1,024$ followed by a linear classifier are used. In contrast to using time-pooled features, this allows for considering the order of frame representations.

\section{Benchmarking Results and Discussion}
\label{sec:results}

Table \ref{tab:DS} shows the full benchmarking results of the various SSL models. The upper and lower parts of the table display the obtained performance for the first and second sets of downstream architectures respectively. The number of neural parameters are also reported both for the SSL encoder and for the downstream decoders. For the latter, only two numbers are provided per task as it only depends on the dimension of the encoder output vector, \textit{i.e.} ``Base'' or ``Large''. In the first set of experiments, the settings of two tasks (LibriSpeech and ASV) are exactly similar to the SUPERB benchmark conditions: the Pearson correlation between our results and those shown in the SUPERB leaderboard reaches respectively 0.99 and 0.97, validating our reproduction of its experimental setting. 

To study the impact of a decoder change on the final performances, we compute, for every task, the Pearson and Spearman correlations between the performances obtained with the first downstream architectures and those obtained with the second ones, and collect them in Table \ref{tab:correlations}. The Pearson correlation evaluates the linear relationship between the two sets of performances, while the Spearman one assesses the strength and direction of their monotonic relationship. Correlation metrics close to $1$ imply proportional performances and similar rankings between the SSL models used with different probes, making the benchmark robust to the considered downstream change.

All the models considered converge to competitive performances on every downstream task and with all the decoding architectures. With the notable exception of LibriSpeech, all the downstream tasks exhibit very different behavior when probes change. In Table \ref{tab:correlations}, the last three columns show the mean performance of the SSL candidates with the first and second downstream decoders. We observe, first, high sensitivity to the chosen decoder, with relative improvements reaching $46.5\%$ and $27.3\%$ respectively for ASV and IC when decoding with the second set of probes. Second, and more interestingly towards answering our main question, the low correlation values (while positive) testify to the large variations in relative performances and rankings between the performances with DS1 and those with DS2. The Spearman correlation, for instance, only reaches $0.34$ and $0.66$ for ER and IC respectively. It is worth noting that while LibriSpeech performance assessment seems robust to changes in the decoder, this does not generalize to the other ASR tasks, starting with the spontaneous English Buckeye corpus with $0.56$ Spearman and $0.42$ Pearson correlation, or the Basque task with $0.19$ Pearson and $0.15$ Spearman correlations. The case of Buckeye ASR is particularly interesting as changing the decoder, from BiLSTM to ContextNet, improves the results for a few models and harms them for others. The best performing one, WavLM Large using the second decoder, is only the fourth best performing one with the SUPERB settings. 

\begin{table}[]
\caption{Pearson and Spearman correlations between the performances obtained with first and second downstream probes, for every considered task. The ``Diff" column shows the relative difference in performance between the two architectures.}

\centering
\scalebox{0.75}{\begin{tabular}{llllll} \toprule
\textbf{Task}   & \textbf{Pearson} & \textbf{Spearman} & \textbf{Mean  DS1} & \textbf{Mean DS2} & \textbf{Diff (\%)} \\ \midrule
LibriSpeech 1-2 & 0.99    & 0.97     & 5.8             & 6.48          & -11.7     \\
Librispeech 1-3 & 0.99    & 0.98     & 5.8             & 7.03          & -21.2  \\ 
Buckeye         & 0.42    & 0.56     & 34.16           & 32.39         & 5.2       \\
Welsh           & 0.59    & 0.62     & 50.64           & 74.52         & -47.2     \\
Basque          & 0.19    & 0.15     & 44.66           & 69.47         & -55.6     \\
VoxCeleb        & 0.47    & 0.75    & 5.2             & 2.78          & 46.5      \\
Iemocap         & 0.22    & 0.34     & 67.66           & 73            & 7.9       \\
Slurp           & 0.75    & 0.66     & 62.1            & 79.04         & 27.3      \\
\bottomrule  
\end{tabular}}
\vspace{-0.7cm}
\label{tab:correlations}
\end{table}

However, we observe that contrarily to the other downstream tasks, the rankings and performances of the considered SSL encoders on the ASR task using LibriSpeech \textit{train-clean-100}, shown in Table \ref{tab:correlations} vary only very lightly with the change of downstream decoder. To cross-validate this finding, experiments using a third downstream decoder, ContextNet, are produced for this task. The results of this additional experiment are reported in Table \ref{tab:DS3}. Similarly, no major differences are witnessed in the ranking of the SSL candidates. For instance, in the three setups, without LM decoding, Data2vec Large is always the best-performing representation, whether it is on the \textit{test-clean} or the \textit{test-other} split. Similarly, DistilHuBERT is systematically the worst performing one and the ``Large" versions of the considered candidates perform consistently better than their ``Base" counterparts. These results are confirmed in Table \ref{tab:correlations} where Spearman and Pearson correlations between the performances using the different downstream decoders are over $0.97$ for LibriSpeech while the highest correlation value is $0.75$ for the other tasks. With LibriSpeech being one of the main ASR benchmarking dataset, and as it is systematically part of the pretraining dataset, these observations suggest that the SSL encoders may be overfitting on this task. These results allow us to conclude that, except for LibriSpeech ASR, current SSL benchmarking is not robust to the choice of the downstream probe.
\begin{table}[]
\centering
\caption{WER results of LibriSpeech experiments on the two considered test splits with a third downstream probe}
\scalebox{0.75}{\begin{tabular}{llllll}

\toprule
\textbf{Tasks \textbackslash Models} & \textbf{SSL  Params} & \textbf{Clean} & \textbf{Other} & \textbf{Clean LM} & \textbf{Other LM} \\ \midrule

DistilHuBERT                         & 23.5M                & 20.52                & 43.27                & 10.44                   & 29.17                   \\
Wav2vec 2.0 Base                     & 95M                  & 7.24                 & 15.66                & 4.73                    & 11.21                   \\
Wav2vec 2.0 Large                    & 317.4M               & 4.35                 & 8.68                 & 03.03                   & 6.86                    \\
HuBERT Base                          & 94.7M                & 7.31                 & 16.00                & 4.60                    & 11.11                   \\
HuBERT Large                         & 316.6M               & 4.04                & 8.63                 & 2.98                    & 6.45                    \\
WavLM Base+                          & 94.7M                & 6.73                 & 15.33                & 4.52                    & 10.84                   \\
WavLM Large                          & 316.6M               & 4.09                & 8.43                 & 2.94                    & 6.15                    \\
Data2vec Base                        & 93.8M                & 5.46                 & 13.34                & 3.76                    & 10.04                   \\
Data2vec Large                       & 314.3M               & \textbf{3.50}                 & \textbf{6.94}                 & \textbf{2.56}                    & \textbf{5.36}                    \\ \midrule
\multicolumn{3}{l}{\textbf{Probe size and inference metrics}} \\ \midrule

\multicolumn{2}{l}{Downstream Parameters Base}                 & \multicolumn{4}{c}{32.4M}         \\
\multicolumn{2}{l}{Downstream Parameters Large}                               & \multicolumn{4}{c}{32.5M}       \\ \bottomrule           
\end{tabular}}
\vspace{-0.2cm}
\label{tab:DS3}
\end{table}
\vspace{-0.02cm}
Furthermore, the reported results highlight an interesting consequence of using low-capacity decoders. With the first set of downstream architectures, ``Large" versions of SSL models are almost always better performing than their ```Base" counterparts. However, this is not the case with higher-capacity decoders in the second set of probes. For instance, the best ASV and ER performances are reached, respectively, with WavLM Base+ and HuBERT Base. For intent classification, when changing the downstream decoder from linear to BiLSTMs, the mean absolute difference between the ``Base" and ``Large" versions' performance drops from $14.23$ to $3.28$. For emotion recognition, while all four ``Large'' versions perform better than the ``Base" ones when probed linearly, enhancing the capacity of the decoder reverses this order for all of them except WavLM. Furthermore, concerning the ASV results, DistilHuBERT performs better with an ECAPA decoder than the best model (WavLM Large) with an x-vector-based head, while containing more than $13$ times fewer parameters. Since the number of parameters does not present a full picture of the involved computations, the THOP library\footnote{\url{github.com/Lyken17/pytorch-OpCounter}} is used to compute the number of Multiply–accumulate operations (MACs). During inference, running DistilHuBERT–ECAPA on VoxCeleb1 enroll and test sets involves computations reaching a mean $249.4$G MACs per batch of $8$ samples, for a result of $2.85$ EER, while WavLM Large–Xvectors requires a mean batch computation of $881.4$G MACs to reach $2.98$ EER. All these observations lead to the conclusion that keeping very small-capacity decoders  may lead to an inflation in model sizes.

\section{Acknowledgements}
This work has benefited from funding from l'Agence de l'Innovation de Défense, and was performed using HPC resources from GENCI-IDRIS (Grant 2021-AD011012801R1).

\section{Conclusion}
In this work, various experiments have been described to evaluate the robustness of speech SSL benchmarks to changes in the downstream probes. The results obtained show a high sensitivity of the leaderboard rankings and relative performances to the choice of the downstream architectures. Furthermore, it is shown that selecting low-capacity decoders may lead to oversized SSL encoders. We hope this diagnosis will help the community to design new benchmarking approaches and encourage submissions to the SUPERB ``Constrained" track.

\bibliographystyle{IEEEtran}
\bibliography{mybib}

\end{document}